# High quality superconducting niobium films produced by Ultra High Vacuum Cathodic Arc

R. Russo[a]

*Dipartimento di Fisica, Università degli studi di Roma " Tor Vergata", 00133 Roma, Italy and*

*Istituto di Cibernetica "E.Caianiello", CNR, 80078 Pozzuoli, Italy*

L. Catani

*INFN-Roma2, Via della ricerca scientifica 1, 00133 Roma, Italy*

A. Cianchi, S. Tazzari

*Dipartimento di Fisica, Università degli studi Roma " Tor Vergata", 00133 Roma, Italy*

*and INFN-Roma2, Via della ricerca scientifica 1, 00133 Roma, Italy*

J. Langner

*The Andrzej Soltan Institute for Nuclear Studies, 05-400 Otwock/Swierk, Poland*

The vacuum arc is a well-known technique to produce coating with enhanced adhesion and film density. Many cathodic arc deposition systems are actually in use in industry and research. They all work under (high) vacuum conditions in which water vapor pressure is an important source of film contamination, especially in the pulsed arc mode of operation. Here we present a Cathodic Arc system working under Ultra High Vacuum conditions (UHVCA). UHVCA has been used to produce ultra-pure niobium films with excellent structural and electrical properties at a deposition temperature lower than $100^{o}C$. The UHVCA technique therefore opens new perspectives for all applications requiring ultra-pure films or, as in the case of Plasma Immersion Ion Implantation, ultra-pure plasmas.

---

[a] Electronic mail: Roberto.Russo@roma2.infn.it

The vacuum arc is one of the oldest technique used for depositing thin films. Its main advantages over other techniques are the ionized state of the evaporated material and the high ion energy,[1,2] allowing deposition of films with enhanced adhesion and higher density.[3] In recent years important progresses have been made in understanding the role of the cathode cohesive energy in the interplay between arc voltage, erosion rate, kinetic ion energy and applied magnetic field[1,4]; the role of potential and kinetic energy in atomic scale heating has been pointed out as well.[5]

The problem of macroparticle production on the cathode surface, affecting the arc sources, has been overcome thanks to the improvement in magnetic filtering,[6] to the point that defect free films can be deposited.[7] The road to new applications where a smooth surface is needed has thus been opened.

To increase the ionization state of the plasma ions and to reduce the thermal load on cathode and substrate several groups developed and used the pulsed vacuum arc technique,[8] which also reduces macroparticle emission. By varying the arc discharge duty cycle and applying a pulsed bias to the substrate the film properties can be further optimized. As an example of the pulsed technique potential, uniform films have been deposited on a substrate with a series of 130 nm wide and ~ 1µm deep parallel trenches.[9]

On the other hand, problems have been reported both in the pulsed and in the DC modes of operation[10] of conventional systems operating in high vacuum (~$10^{-7}$ hpa), such as incorporation of oxygen and hydrogen in metal films.[11] The problem is believed to originate from the water vapor layer that forms on the cathode surface in between arc pulses because of the residual water vapor present in the chamber. The problem can be solved by lowering the pressure to UHV standards, thus opening the road towards applications requiring ultra-pure metallic films.

The only UHV arc reported in literature to our knowledge[12] has been operated at 600V and only 150mA, below the parameter space of a true cathodic arc (a few tens of Volts at currents in the $10^2$ A range). We present here a truly UHV cathodic arc (UHVCA) system designed mainly in view of application to niobium coating of superconducting RF cavities for particle accelerators.[13] It is in fact expected that Nb arc–coated Cu cavities would allow reaching useful accelerating fields higher than the present ~15MV/m of cylindrical magnetron sputtered ones.[14]

A schematic drawing of the systems is shown in Fig.1. It is pumped down by an oil-free pumping system consisting of a membrane pump on the foreline of a turbo molecular drag pump and reaches a base pressure of ~1x10$^{-10}$ hPa after a 12h bake-out at 150°C. Ignition of the arc at such a low pressure is not only more critical than for a standard arc device but the triggering system must be absolutely "clean" not to contaminate the plasma. A common triggering method that uses high voltage discharge across the surface of an insulator had to be discarded because traces of elements evaporated from the insulator were found in films, particularly when several ignitions were needed during a single deposition. After testing several other methods[15] a laser triggering system[16] has been finally adopted. A compact 50 mJ Nd-YAG laser focused on the cathode is sufficient to most reliably ignite the arc under the cleanest possible conditions.[b] To prevent contamination due to memory effects when a same systems is used to deposit different materials[18] we have only used high purity niobium cathodes.

---

[a] More details on the system can be found in ref. 17

Vacuum condition before and during the discharge are checked using a Residual Gas Analyzer (RGA). Immediately after the arc is started, outgassing of light elements (mainly $H_2$, CO, hydrocarbides and water) is observed but after ~1 minute of operation the partial pressure of all outgassed species other than hydrogen, directly coming from the Nb cathode, fall to below ~$10^{-9}$ hPa. The total pressure during DC operation is typically in the $10^{-7}$ hPa range, 99% of it being due to hydrogen (see Fig.2). As soon as the arc extinguishes the hydrogen pressure drops one order of magnitude, reaching its equilibrium pressure in presence of a fresh oxide-free Nb surface, then starts slowly decreasing pumped by the 200 l/s turbomolecular pump.

Niobium films have been deposited on both copper and sapphire substrates. Samples were placed on an electrically insulated sample holder, at about 50cm from the cathode surface. A constant 40V bias was applied to holder so as to deflect electrons and collect only the multiply charged, >100eV kinetic energy Nb ions.[1] The thermal load on the sample was thus much reduced and a thermocouple placed on the sample holder indicated that the substrate temperature was kept below 100°C during the deposition. The average kinetic energy of the Nb atoms, reaching the substrate with an average charge of +3, is estimated to be ≈250eV, a value large enough to produce some sputtering from the growing film. The lowest arc current for stable operation was about 60A and its maximum value, limited by our cooling power, was 160A in the DC mode.

Samples were produced at several arc current values and their electronic and structural properties analyzed. In particular, inductive measurements of the superconducting critical temperature $T_c$ (9.26K for pure bulk Nb), very sensitive to impurity and stresses, were carried out to have a first indication of the film quality. It is for instance known that very small amounts of oxygen in the film can lower $T_c$[19] while compressive stresses can raise it up to 9.6K.[20]

Inductive measurements were carried out by placing the sample in contact with a primary coil producing a 1KHz oscillating e.m. field. The sample temperature is then varied looking for response on the third harmonic of the exciting frequency, only present during the transition to the superconducting state.[21] Typical results are shown in Fig.3 for 6 samples. Differences in $T_c$ compared to the bulk value (bulk $T_c$ = 9.26K) are small and can be in part due to small temperature differences between the thermometer and the samples. No dependence on either film thickness, in the $0.1 \div 2$ $\mu m$ range, or arc current were found, all samples being high quality. The sharp transition widths (<0.02K), close to that of bulk Nb, are strong indicators of uniform and clean films. The absolute $T_c$ value indicates that stresses in our films are lower than in magnetron sputtered films, an indication confirmed by the XRD analysis (in a $\theta/2\theta$ configuration) yielding a lattice parameter in the $0.3308 \div 0.3318$ nm range, a value close to the 0.3306 nm of bulk Nb.

Confirmation of the above measurements is also provided by $T_c$ measurements of samples with an important fraction of the surface covered with macroparticles[c], molten Nb drops, 0.1÷10μm in diameter, that can be assumed to behave like bulk material. In fact the inductive measurement reveals two separate transitions, the film and the bulk one, as shown in Fig. 4. The transition curve shapes are very similar and, because the coupling to the coil is the same, the step height is proportional to the percentage of covered surface.

Another parameter very sensitive to impurity is the Residual Resistivity Ratio (RRR), defined as the room temperature resistivity divided by that at 10K: the higher the RRR, the higher the purity. Typical RRR values of Nb films deposited by sputtering on room temperature substrates range from 2 to 10, while bulk Nb with RRR in the $40 \div 500$ range is commercially available. RRR values measured on our Nb film samples, deposited on sapphire at room temperature, range from 20 to 50, while heating the substrate at 150$^\circ$C resulted in niobium films with RRR around 80. RRR values up to 40 can also be obtained by magnetron sputtering in UHV systems but only at substrate temperatures higher than 200$^\circ$C, not compatible with all applications (e.g: copper substrate is annealed and its mechanical properties degraded; in multilayers deposition diffusion in between layers may occur). We believe that such high value of RRR could be obtained in our film thanks to the atomic scale heating due to the kinetic and potential energy of the niobium ions reaching the films surface, resulting in a local temperature higher than average temperature reached by the substrate and recorded by the thermocouple.

---

[c] Details on macroparticle distributions and filtering are reported elsewhere[17].

In conclusion we have produced very high quality superconducting Nb film samples at room temperature in what we believe is the first cathodic arc system working in ultra high vacuum. The UHVCA system has shown excellent reproducibility and reliability also thanks to laser triggering. It is ultra-clean and after ~ 1 minute conditioning the partial pressure of all outgassed species other than hydrogen, fall to below ~$10^{-9}$ hPa. The total pressure during arc operation is in the $10^{-7}$ hPa range, 99% of it being due to hydrogen outgassed from the niobium cathode. The UHVCA opens the road towards applications requiring ultra-pure metallic films, in particular on substrate that cannot be heated at high temperatures, or ultra-pure metallic plasma for plasma immersion ion implantation.

Figure captions

Fig. 1) Schematic drawing of the UHVCA system. The cathode(1) is mounted on a water cooled copper support (2) and inserted in a water cooled stainless steel anode (3). A niobium floating potential electrode(4) prevents the discharge from moving to the bottom part of the system. The magnetic coil (5) confines the hot spot on the flat cathode surface, while magnetic Helmotz coils (6) improve the deposition rate on the sample holder (7). The triggering is provided by a laser beam focused on the cathode through an optical window (8)

Fig. 2) Behavior of several ion current versus time as recorded by RGA. When the arc starts all ion currents increase, but after about one minute operation the ion current of all gases become more than 2 orders of magnitude lower than the hydrogen one. The most important contaminants (water, CO and light hydrocarbides) are found in $10^{-9}$hPa range and their partial pressures decrease with time of arc operation, while hydrogen ion current is constant to about $2\times10^{-7}$hPa partial pressure.

Fig. 3) Inductive $T_c$ measurement for six typical niobium samples produced by UHVCA. The film thickness ranges from 100nm to 2μm. The sharp transition width (<0.02K) is similar to that of bulk and gives a strong indication of uniform and clean films. The absolute $T_c$ values indicate that stresses in the films are lower than in the magnetron sputtering case.

Fig. 4) Inductive $T_c$ measurements for two Nb samples with macroparticles covering a large fraction of the surface. For each sample two transitions are visible: the first due to the film and the second due to macroparticles. In both cases transition widths are about 0.01K for both film and bulk. Small differences in the macroparticle $T_c$ are within the instrument reproducibility error (about 0.03K).

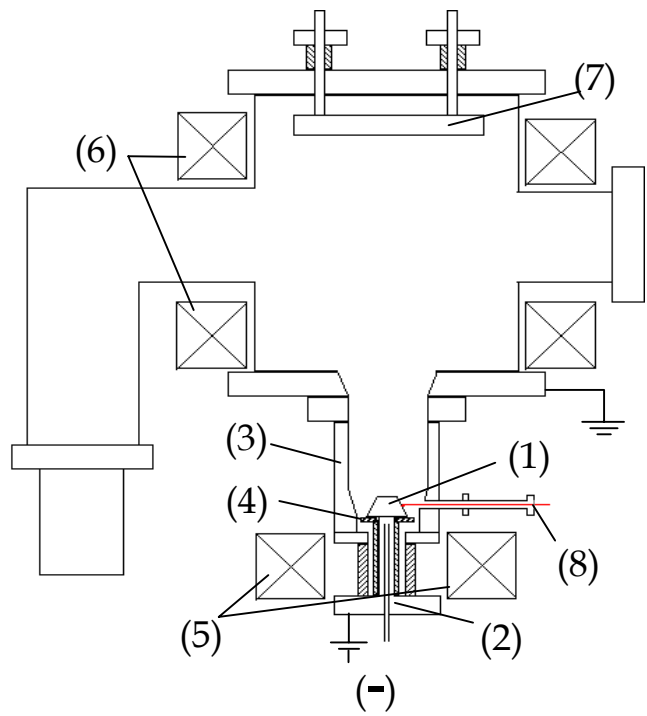

R. Russo et al.  Fig.1

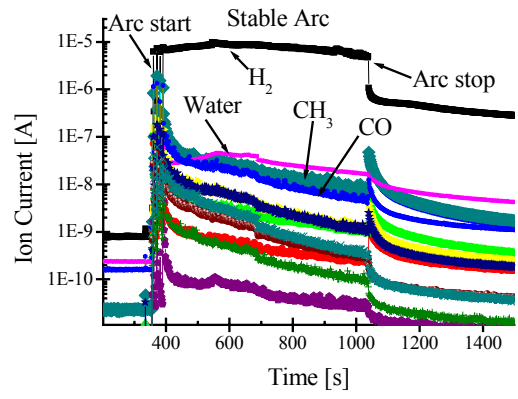

R. Russo et al. Fig.2

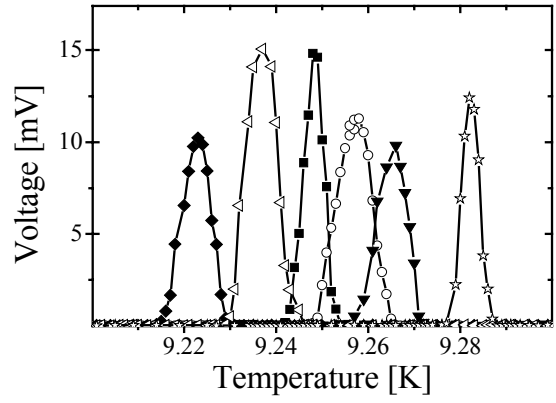

R. Russo et al.  Fig. 3

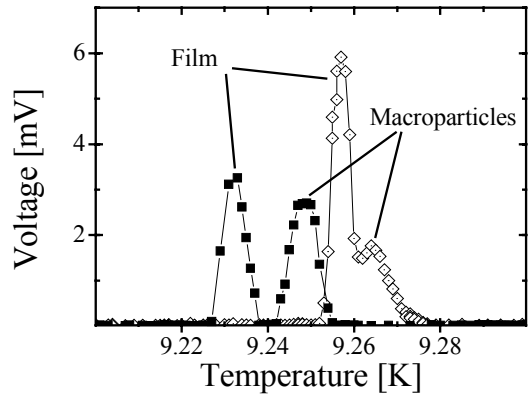

R. Russo et al.  Fig. 4